# Phosphides-based terahertz quantum-cascade laser

D. V. Ushakov,[1] A. A. Afonenko,[1] R. A. Khabibullin,[2,3] M. A. Fadeev,[4] A. A. Dubinov[4,5,*]

[1]*Belarusian State University, 4 Nezavisimosti Avenue, Minsk, 220030, Belarus*
[2]*V.G. Mokerov Institute of Ultra-High Frequency Semiconductor Electronics RAS, 7/5 Nagornyy Proezd, Moscow, 117105, Russia*
[3]*Moscow Institute of Physics and Technology, 9 Institutsky Lane, Dolgoprudny, 141701, Russia*
[4]*Institute for Physics of Microstructures, Russian Academy of Sciences, GSP-105, Nizhny Novgorod, 603950, Russia*
[5]*Lobachevsky State University of Nizhny Novgorod, 23 Gagarina Avenue, Nizhny Novgorod, 603950, Russia*
*sanya@ipmras.ru

**Abstract**

Due to their high optical phonon energies GaInP/AlGaInP heterostructures are a promising active medium to solve the problem of creating compact semiconductor sources with an operating frequency range of 5.5-7 THz. In this work, the temperature dependences of gain and absorption at 6.8 THz have been calculated for a GaInP/AlGaInP-based quantum-cascade laser (QCL) with two quantum wells in the cascade and a metal-metal waveguide. We propose a laser structure that provides a mode gain of over 90 cm$^{-1}$ with a maximum operating temperature of 104 K. The results of this study open the way to the development of a QCL for operation in a significant part of the GaAs phonon absorption band region, which is inaccessible for existing QCLs.

Since their invention two decades ago THz quantum cascade lasers (QCLs) have progressed from cryogenic devices with relatively small output power [1] to powerful THz sources operating under thermoelectric Peltier cooling [2, 3]. However, QCLs are not yet available in the range from 5.5 to 10.5 THz despite its importance in such applications as spectroscopy of organic and non-organic materials, liquids and gases [4-7]. The main obstacle for making QCLs for this range is a strong phonon absorption in the materials that they are made of (GaAs/AlGaAs, InGaAs/InAlAs) [8, 9]. Therefore, semiconductors with a different energy of polar-optical phonons can serve as an alternative to group III metal arsenides for the creation of QCLs. Previously, both $A_{III}B_V$ materials (GaN/AlGaN [10]) and $A_{II}B_{VI}$ materials (ZnO/MgZnO [11], ZnSe/ZnMgSe [12], HgCdTe [13]) were suggested as such semiconductors. It has also been suggested using van der Waals heterostructures of 2D - materials with graphene [14]. However, the experimental realization of QCLs based on the proposed materials showed only spontaneous electroluminescence due to the complexity of precise growth of hundreds of nanoscale layers of these materials (see, for example, works on GaN/AlGaN [15], ZnO/MgZnO [16]) or difficulty to form ideal structures based on 2D materials [17].

In this paper we investigate the prospects of using a material system based on group III metal phosphides (GaInP/AlGaInP) for creating a double metal waveguide QCL with generation frequency of ~ 7 THz. The choice of this heteropair is explained by the fact that GaInP and AlGaInP are polar semiconductors with optical phonon frequencies in the range of 10 - 12 THz, which is higher than that of the traditional GaAs/AlGaAs heteropair (phonon energy ~ 8 THz) [18]. A major advantage of phosphide-based heterostructures is the developed growth technology using both molecular beam epitaxy and vapor deposition. Similarly to GaAs, GaInP and AlGaInP have a zinc blende lattice structure. The lattice constants of $Ga_{0.51}In_{0.49}P/(Al_xGa_{1-x})_{0.51}In_{0.49}P/GaAs$ are very close, which allows the growth of relaxed multilayer structures on GaAs substrates. In addition, quite a lot of work has been devoted to the growth of high-quality GaInP/AlGaInP superlattices, which were used to create red-orange interband lasers (see, for example, [19-21]).

To model the THz QCL based on GaInP/AlGaInP, we used a model based on a system of balance equations for localized states and continuum states. To account for the effects of dephasing on the charge transport processes, we employed the technique of modifying the eigenbasis of Schrödinger equation by reducing the dipole moments of tunnel-bound states. The algorithm for calculating optoelectronic properties includes determining the energy levels and wave functions based on the solution of Schrödinger equation within 3-band **kp**-approximation, calculating matrix elements of dipole transitions, calculating scattering rates on optical phonons, ionized impurities, roughness of hetero-boundaries and an electron-electron scattering, determining surface concentrations of carriers for the corresponding energy subbands from the closed system of balance equations, calculating the electric current and amplification spectra. The details of the calculation method tested on GaAs/AlGaAs heterostructure and a comparison of the calculations with experimental results showing good agreement can be found in [22, 23].

For our calculations we chose with the design with two quantum wells (QWs) in a cascade and a resonant-phonon depopulation scheme, when the operating voltage multiplied by the electron charge $q$ in one cascade is close to the sum of the energies of the radiated photon and the longitudinal optical phonon. Our choice of the number of QWs in a cascade is justified by the operating temperatures of THz QCLs based on GaAs/AlGaAs with two-QW design being the highest [3]. In the course of optimization (scanning the thicknesses of the layers of the structure to obtain the highest gain at 6.8 THz), we found the following sequence of layers in one cascade in nm: **4.8**/5.36/**1.98**/12.98 (**17**/19/**7**/46 monolayers of the corresponding semiconductor) with $(Al_{0.5}Ga_{0.5})_{0.51}In_{0.49}P$ barriers (in bold) and $Ga_{0.51}In_{0.49}P$ QWs (see Fig. 1). The central part of the underlined QW is assumed to be doped with a layer electron concentration of $4.64 \times 10^{10}$ cm$^{-2}$. At the operation bias point $V_1 = 73$ mV per cascade, injector levels $i'$ and upper laser level $u$ are aligned which leads to the resonant tunneling of electrons. The transition of an electron from the upper laser level $u$ to the lower laser levels $l$ is accompanied by the diagonal THz photon emission ($\nu_{ul} = 6.8$ THz) with dipole matrix element $Z_{ul} = 2.44$ nm and high population inversion of approximately 42%. Finally, electrons in the lower lasing levels relax to the injector level $i$ through fast electron–longitudinal phonon scattering. This process then repeats in the subsequent cascades.

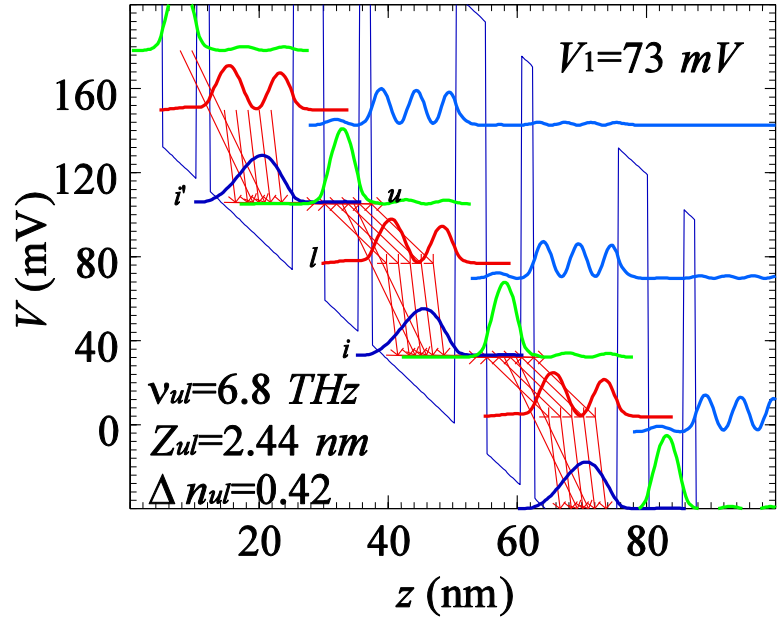

*Fig. 1. Energy profile of the bottom of the conduction band V, energy levels and squares of the moduli of the wave functions in the growth direction of the proposed structure at the value of the applied voltage $V_1 = 73$ mV per cascade and temperature $T = 77$ K. The number of arrows is proportional to the current density.*

In order to estimate the maximum operation temperatures of the given designs, we calculate the dielectric function and cavity losses for a 10 μm thick Au/Ti-Ti/Au waveguide, based on InGaP, as a function of temperature using the method proposed in Ref. [24]. This calculation accounts for the losses in metal layers, on optical phonons and free charge carriers.

The dielectric constant of metals in the THz spectral region is usually described according to the Drude model as

$$\varepsilon_m(\nu) = 1 - \frac{\nu_{pm}^2}{\nu(\nu + i\gamma_{pm})}, \qquad (1)$$

where $\nu_{pm} = (Nq^2/m_e\varepsilon_0)^{1/2}$ is the plasma frequency, $N$ is the free-electron concentration, $m_e$ is the effective electron mass, $\varepsilon_0$ is the vacuum permittivity, and $\gamma_{pm}$ is the Drude damping parameter. The temperature dependence of the damping parameter was determined through its linear relationship with the temperature dependence of the resistivity of the metal [24]. The values of $\nu_{pm}$ and $\gamma_{pm}$ were calculated to be 9.02 eV and 12.3 meV (80 K) for Au, which agree well with the data of [24, 25]. We are taking into account absorption in the Ti (5 nm) adhesion layer with parameters $\nu_{pm}$=8.84 eV, $\gamma_{pm}$=323.3 meV (80 K) [26, 27]. This layer prevents Au diffusion into the epitaxial layers [2].

The active region of the QCL is sandwiched between two $n^+$-InGaP contact layers with a dopant concentration of $10^{18}$ cm$^{-3}$ (27-nm thick) to form ohmic contacts. The complex dielectric function of semiconductor layers was constructed as a superposition of three damped harmonic oscillators corresponding to the AC- and BC-like phonon modes for ternary alloys $A_{1-x}B_xC$ and to the free carriers, respectively. Hence [28]:

$$\varepsilon(\nu) = \varepsilon_\infty \prod_{j=1}^{N} \frac{\nu^2 - \nu_{LO,j}^2 + i\nu\gamma_{LO,j}}{\nu^2 - \nu_{TO,j}^2 + i\nu\gamma_{TO,j}} - \frac{\nu_{ps}^2}{\nu(\nu + i\gamma_{ps})}, \qquad (2)$$

where $\varepsilon_\infty$, $\nu_{LO,j}$, $\nu_{TO,j}$, $\gamma_{LO,j}$, $\gamma_{TO,j}$, $\nu_{ps}$, $\gamma_{ps}$ represent, in order, the high frequency dielectric constant, the LO-,TO- phonon frequencies and the corresponding phenomenological damping parameters, semiconductor plasma frequency and the damping parameters of free carriers. The temperature dependence of the damping parameter of semiconductor layers was calculated with the help of the expression: $\gamma_{ps}(T) = q/m_e\mu(T)$, where $\mu(T)$ is the electron mobility as a function of temperature. The dependences of the mobility on temperature and impurity concentration for QCL layers were found on the basis of experimental data using interpolation formulas according to [29]. The values $m_e$= 0.0987 free electron mass, $\mu_{QW}$=4207 cm$^2$/Vc, $\mu_{n+}$= 800 cm$^2$/Vc common to both In$_{0.49}$Ga$_{0.51}$P doped quantum wells and n$^+$-layers at the 80 K.

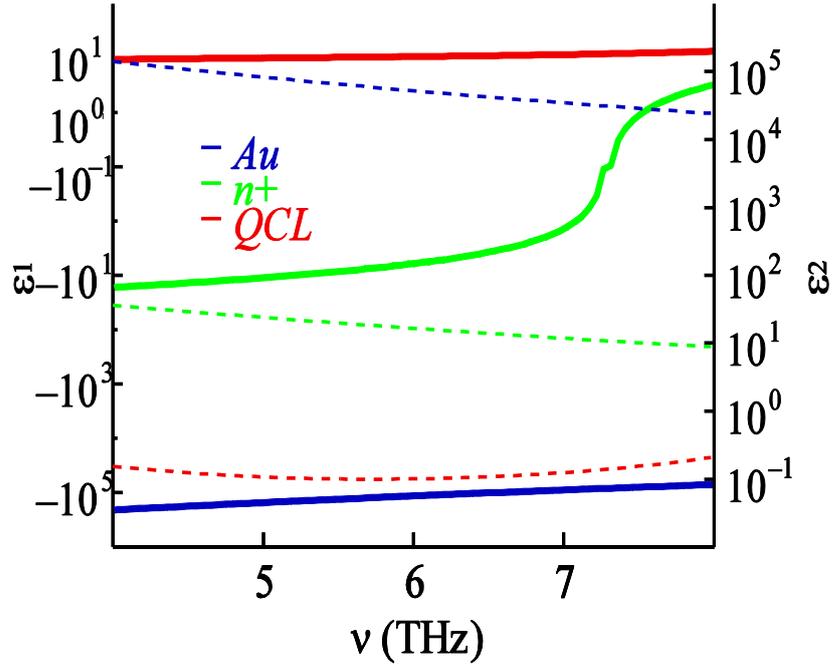

*Fig. 2. Real ($\varepsilon_1$) and imaginary ($\varepsilon_2$) part of the dielectric function spectra for the main components of the QCL waveguide. T = 80 K.*

The main contribution to losses of the active region and n$^+$ layers is given by the absorption in In$_{0.49}$Ga$_{0.51}$P layers. The highfrequency dielectric function, GaP(1)- and InP(2)-like mode frequencies and dumping parameters (in cm$^{-1}$) for In$_x$Ga$_{1-x}$P are approximately given by [28, 30, 31]:

$$\varepsilon_\infty = 9.11(1-x) + 9.61x; \qquad (3)$$

$$\begin{aligned}
\nu_{LO1} &= 404.99 - 38.97x - 18.18x^2, \\
\nu_{TO1} &= 395.02 - 54.26x + 6.72x^2, \\
\gamma_{LO1} &= 3.4\frac{T}{300}, \gamma_{TO1} = 8.1\frac{T}{300}.
\end{aligned} \qquad (4)$$

$$\begin{aligned}
\nu_{LO2} &= 394.59 - 80.36x + 30.26x^2, \\
\nu_{TO2} &= 368 - 88.95x + 26.04x^2, \\
\gamma_{LO2} &= 9.5\frac{T}{300}, \gamma_{TO2} = 9.92\frac{T}{300}.
\end{aligned} \qquad (5)$$

In order to obtain the value of dielectric function of the quaternary alloy (Al$_{0.5}$Ga$_{0.5}$)$_{0.51}$In$_{0.49}$P for the barriers we calculated the average of ternary alloys In$_{0.49}$Ga$_{0.51}$P and Al$_{0.52}$In$_{0.48}$P. The high frequency dielectric function, InP(1)- and AlP(2)-like mode frequencies and dumping parameters for Al$_x$In$_{1-x}$P are approximately given by [28, 32]:

$$\varepsilon_\infty = 7.53x + 9.61(1-x) - 1.72x(1-x); \qquad (6)$$

$$\nu_{LO1} = 350.8x + 344.5(1-x),$$
$$\nu_{TO1} = 303.3x + 345.9(1-x), \quad (7)$$
$$\gamma_{LO1} = 21\frac{T}{300}, \gamma_{TO1} = 12.7\frac{T}{300}.$$

$$\nu_{LO2} = 501x + 412.4(1-x),$$
$$\nu_{TO2} = 440x + 414.3(1-x), \quad (8)$$
$$\gamma_{LO2} = 5.1\frac{T}{300}, \gamma_{TO2} = 31.3\frac{T}{300}.$$

The dielectric function spectra for the main components of the QCL waveguide is shown in Fig. 2.

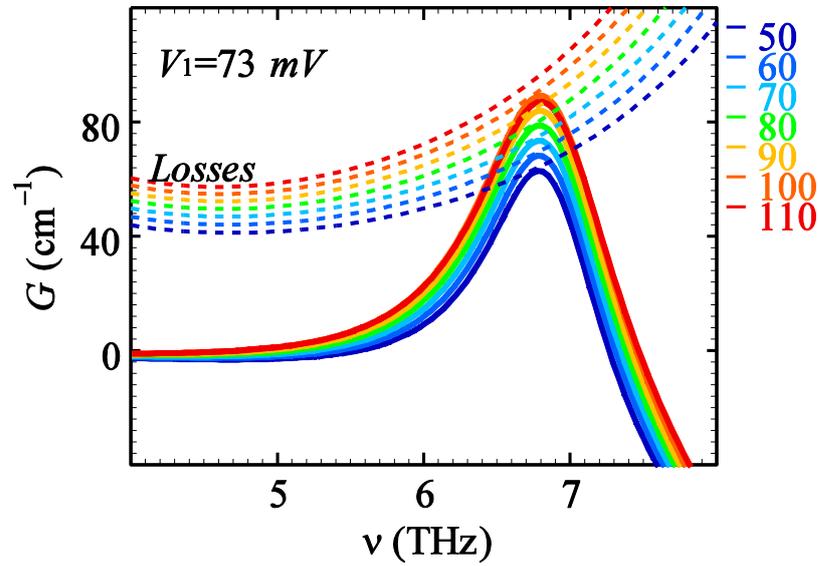

*Fig. 3. The spectra of gain (solid curves) and the losses (dashed curves) in the generation mode for different temperatures.*

Calculations of waveguide characteristics were performed for the TM-mode through numerical solution of the wave equation [24]. Calculations of the temperature transformation of the loss and gain spectra in the generation mode, when the gain is equal to the loss coefficient, are shown in Figure 3. The coefficient of total losses for temperatures 40, 60, 80, 100 K takes values: 53.3, 70.2, 80.8 and 91.6 cm$^{-1}$ respectively. the calculations show. Contributions to the coefficient from absorption in metal layers, phonon absorption and radiation absorption by free carriers for temperature of 80 K and frequency of 6.8 THz are: 19.8, 37.2 and 22.8 cm$^{-1}$ respectively. Reflection losses on the mirrors were taken into account in addition to the waveguide mode losses. For a 1-mm long resonator with reflection coefficient R = 0.9, the reflection losses amounted to 1 cm$^{-1}$. The maximum of the gain spectrum corresponds to diagonal laser transition *u-l*, which is realized at ~ 6.8 THz as can be seen in Fig. 3. The calculations performed for the proposed QCL show that the gain at 6.8 THz exceeds 90 cm$^{-1}$, which would allow such a laser to operate up to a sufficiently high temperature of 104 K (Fig. 4). At that, the generation frequency slightly decreases with temperature but remaining in the region of 6.7-6.8 THz.

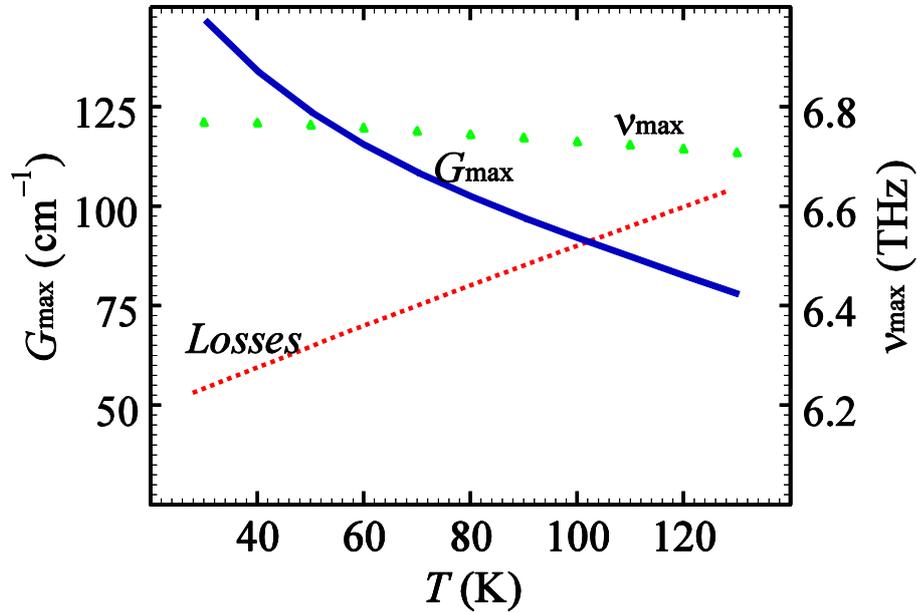

*Fig. 4. Temperature dependence of the maximum gain, corresponding losses and generation frequency $\nu_{max}$. $V_1$= 73 mV.*

In conclusion, we summarize the main results of the work. We carried out the modeling of the temperature characteristics of a 6.8 THz QCL based on GaInP/AlGaInP with a two-QW cascade design and a metal-metal waveguide. The results show the possibility of finding the optimal GaInP/AlGaInP materials and layer thicknesses in order to obtain lasing at temperatures up to 104 K. We believe that GaInP/AlGaInP structures can be promising candidates for the QCLs operating in the frequency range of 5.5-7 THz.

**Acknowledgements**
The work was supported by the Russian Science Foundation, grant # 23-19-00436, https://rscf.ru/project/23-19-00436/.

**Disclosures**
The authors declare no conflicts of interest.

**Data availability**
The data that support the findings of this study are available from the corresponding author upon reasonable request.